\documentclass[useAMS,usenatbib]{mn2e}
\usepackage{graphicx,amsmath,natbib}
\usepackage{times}

\title[Cooling of Quark Stars in the 2SC Phase]{Cooling of Quark Stars in the Color Superconductive Phase:
{\it Effect of Photons from Glueball decay}}

\author[Ouyed et al.]{Rachid Ouyed$^{1,2}$, M. J. Hamp$^{3}$, and S. C. Woodworth$^{3}$\\
$^{1}$ Department of Physics and Astronomy, University of Calgary,
       2500 University Drive NW, Calgary, Alberta, T2N 1N4 Canada\\
$^{2}$ Canadian Institute for Theoretical Astrophysics, 60 St.
 George Street, Toronto, ON M5S 1A7, Canada\\
$^{3}$ Department of Engineering Physics, McMaster University,
Hamilton, Canada}

\date{Released 2005 Xxxxx XX}

\pagerange{\pageref{firstpage}--\pageref{lastpage}} \pubyear{2005}

\def\LaTeX{L\kern-.36em\raise.3ex\hbox{a}\kern-.15em
    T\kern-.1667em\lower.7ex\hbox{E}\kern-.125emX}

\begin{document}

\label{firstpage}

\maketitle

\begin{abstract}
The cooling history of a quark star in the color
 superconductive phase is investigated. Here we specifically
focus on the 2-flavour color (2SC) phase where novel
process of photon generation via glueball (GLB) decay have been already
 investigated (Ouyed \& Sannino 2001).
 The picture we present here can in principle be generalized to quark stars entering a superconductive phase where similar photon generation mechanisms are at play. 
 As much as $10^{45}-10^{47}$ erg of energy is provided
by the GLB decay in the 2SC phase.  
 The generated photons  slowly diffuse out of the quark star keeping it  
hot and radiating as a black-body (with possibly a Wien spectrum in gamma-rays) for millions of years. We discuss hot radio-quiet isolated neutron stars
 in our picture (such as  RX J185635-3754 and RX J0720.4-3125)  and argue that 
  their nearly blackbody spectra (with a few broad features) and their
 remarkably tiny hydrogen atmosphere are indications
 that these might be quark stars in the color superconductive phase where some sort of photon generation mechanism (reminiscent of the GLB decay) has taken place. 
Fits to observed data of cooling compact stars favor models with superconductive 
gaps of $\Delta_{\rm 2SC} \sim 15$-$35$ MeV 
  and densities $\rho_{\rm 2SC}=(2.5$-$3.0)\times
\rho_{\rm N}$ ($\rho_{\rm N}$ being the nuclear
matter saturation density) for quark matter in the 2SC phase.
If correct, our model combined with more observations
of isolated compact stars could 
provide vital information  to studies of quark matter and its exotic phases.
\end{abstract}

\section{Introduction}

Quark matter at very high density  is
expected to behave as a color superconductor (e.g. Rajagopal \& Wilczek 2001).
Of interest to the present work is the 2-flavor color
superconductive phase, the 2SC phase,  where only the 
up and down quarks of two color (say, $u_1,u_2,d_1$, and $d_2$)
 are paired in a single condensate, while the ones of the
third color (say, $u_3$ and $d_3$) and the strange quarks
of all three colors are unpaired.
 Associated with superconductivity is the so-called gap energy $\Delta_{\rm 2SC}$
inducing the quark-quark pairing and the critical temperature ($T_{\rm c}$)
above which thermal fluctuations will wash out the superconductive state.

The cooling of quark stars in such a  
phase\footnote{Here we only consider the
 stars in 2SC phase. Cooling of quark stars in the Color-Flavor-Locked
phase where the strange quark enters the dynamics,  has   
also been investigated in the literature
and it was shown they cool down too rapidly in disagreement with
observations (e.g., Schaab et al. 1997; Blaschke et al. 2001).}
 has previously been investigated in the literature.
It was found that a crust must be included for 2SC star cooling
to be compatible with existing data (Blaschke et al. 2000).
As for the cooling of hybrid stars (neutron stars
with a 2SC core), it is believed that 
small gaps ($\Delta_{\rm 2SC} < 1$ MeV) tend to reproduce cooling
 curves which agree well
with the observed data (we refer
the interested reader to \S 6 in Weber 2005 and references therein). 
Whether such small gaps can be justified 
and why the crust - believed to be tiny in quark
stars - is so crucial to cooling is debatable.  More recent studies
 of these hybrid stars find that unless all quarks
 are gaped the stars cool too fast in disagreement with
observational data (the fast cooling is due to direct Urca process
 on unpaired quarks; we refer the reader to Blaschke et al. 2005
 for more details).  However the studies mentioned
 above did not include internal heating mechanisms
 as we do here (the GLB decay) which as we will
show have definite consequences on the cooling history
 and spectral features of these stars.
 Furthermore, there are still pulsar candidates these models fail
to explain leaving room for exploring other possibilities.

A novel feature of the 2SC phase is the generation of GLB particles
(hadrons made of gluons) which as demonstrated
in Ouyed \& Sannino (2001) immediately decay into photons. 
 These GLBs appear at temperatures much  
below the critical temperature for the onset of the superconductive phase.
 We isolate three cooling steps:
 (i) in the early hot stages when 
 the quarks are still unpaired  the neutrino emission
is due to the three quark direct Urca processes $u_c+e^-\rightarrow
d_c+\nu_e$ and $d_c\rightarrow u_c+e^-+\bar{\nu}_e$, 
where the subscript $c$ denotes the quark
color. The star rapidly cools to $T_{\rm c}$
thus entering the 2SC phase; (ii) in the  2SC phase  
 the paired up and down quarks of two colors do not contribute to cooling thus 
reducing neutrino emission (see Page\&Usov 2002 for more details).
 While reduced by a factor of $3$, the neutrino emission remains
the dominant cooling mechanism until the star has cooled
to the critical temperature for GLB formation;
 (iii) at this point in time, the GLBs decay to photons providing an extra heat source
 and the cooling becomes driven by these photons slowly escaping from the star.
 We should note that this photon cooling stage might be
particularly dominant for the further cooling history of the star 
  for cases and situations where neutrino cooling is  
quenched (e.g., by secondary pairing of 
the $u_3$ and $d_3$ quarks as discussed in Page\&Usov 2002).  
This third phase, unique to our model, alters 
the cooling history of the star keeping hot and
radiating as a blackbody for millions of years as we show in this work.

\subsection{Caveats on the 2SC phase and model generalities}

It has been argued that once the conditions of charge neutrality
and $\beta$-equilibrium are enforced it appears
that the so-called ``gapless 2SC phase", or g2SC, might be the favoured
ground state  (Shovkovy \& Huang 2003;  Aguilera et al. 1995).
Other studies instead favor the so-called ``gapless CFL phase", or gCFL
(Alford, Kouvaris, \& Rajagopal 2004; and references therein).
These alternatives (g2SC and gCFL) however might be prone to
instabilities begging for serious scrutiny before this debate
can be settled (e.g. Shovkovy, R\"uster, \& Rischke 2004; Huang \& Shovkovy
 2004; He et al. 2005).
The differences in these studies
are connected with the choice of the model parameters
and is beyond the scope of this paper.
For g-2SC matter the most likely resolution of the instability 
is a transition to a crystalline superconductor (e.g. 
 Casalbuoni et al. 2005)\footnote{It remains to be shown 
 that  such a phase possesses a thermodynamic potential (free energy)
which is below the unstable g-2SC phase.   While this might be the case
 in a narrow density range in the QCD phase space,
 more studies involving more complicated crystalline structures
seem to be  needed to settle this issue (Casalbuoni 2003).}.  
If such a transition occurs on much longer timescales
than GLB formation and decay time (i.e. $10^{-13}$-$10^{-14}$ s; Ouyed \& Sannino 2001) 
then our model should still be valid;  a stable 2SC/g-2SC phase is not
  a necessary condition.  
  As we have said, the picture we are presenting here in its generality  should 
 apply to quark stars in any phase where photon generation mechanisms are at
play (e.g. Vogt, Rapp, \& Ouyed 2004).

The paper is presented as follows: In Sect. 2 we
briefly present the GLB formation
and decay in the 2SC phase followed
in Sect. 3 by a description of the model's
assumption. Cooling calculation
results are shown in Sect. 4 with a comparison
to observed data. In particular, radio-quiet isolated neutron stars  
  are discussed in the
context of our model in Sect. 5. We conclude in Sect. 6.

\section{Glueball formation and decay}

The GLB mass is given in Ouyed\&Sannino (2001) as,
\begin{equation}
M_{\rm GLB}^2 = \frac{\sqrt{b}}{2d\sqrt{e}} \hat{\Lambda}^2\ , 
\end{equation}
where $b=22/3$, $e^2/4\pi=1/137$ and $d$ is a positive
constant of order unity, and 
\begin{equation}
\hat{\Lambda}\simeq \Delta \exp\left( -\frac{2\sqrt{2}\pi}{11}\frac{\mu}{g(\mu)\Delta}\right)\ .
\end{equation}
In the expression above,  
$\mu \propto \rho^{1/3}$ is the chemical potential ($\rho$ is the quark matter density) and $g(\mu)$
the effective energy dependent coupling constant
define as $g(\mu)^2/4\pi\sim 1/\ln(E/\Lambda_{\rm QCD})$
where $\Lambda_{\rm QCD}$ is the QCD energy scale.
Here $E$ is the energy scale at which the theory is being applied
which in our case is $E\sim \mu$.

\begin{figure}
\includegraphics[width=0.5\textwidth]{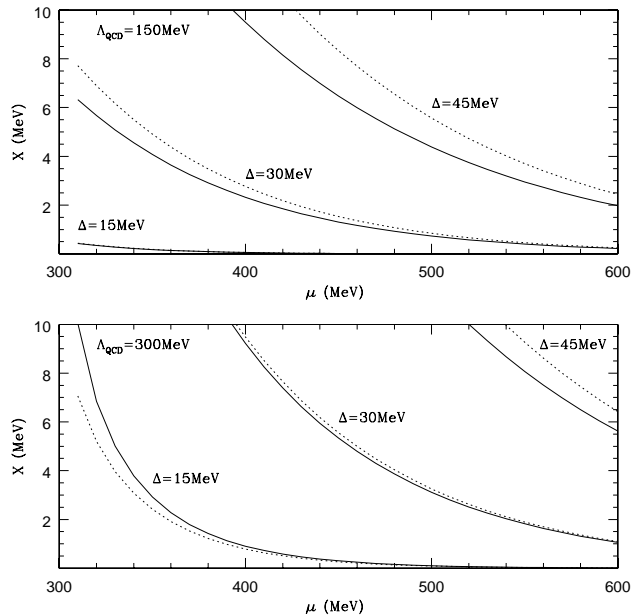}
\caption{Glueball mass (solid) and
deconfinement temperature (dotted) vs chemical potential 
for two different QCD energy scales, $\Lambda_{\rm QCD}=150$ MeV (top panel)
and  $\Lambda_{\rm QCD}=300$ MeV (lower panel).
 }
\label{fig:gbmass}
\end{figure}

It has been shown in Sannino et al. (2002) that the
GLBs  appear (or enter the dynamics)
 at the critical temperature,
\begin{equation}
T_{\rm GLB} = \sqrt[4]{\frac{90v^3}{2e\pi^2}}\hat{\Lambda}\ ,
\end{equation}
 (for $T>T_{\rm GLB}$, GLBs ``melt") 
where $v=1/\sqrt{\lambda \epsilon}$ is the gluon velocity
with the dielectric constant ($\epsilon$) and the magnetic
permeability ($\lambda$) given as (Rischke et al. 2001)
\begin{equation}
\epsilon = 1 + \frac{g(\mu)^2\mu^2}{18\pi\Delta^2}\ , \qquad \lambda=1
\end{equation}

Figure~\ref{fig:gbmass} shows the mass, $M_{\rm GLB}$, and $T_{\rm GLB}$  
of the GLBs for different gaps and chemical potentials
for the two cases of $\Lambda_{\rm QCD}=150$ MeV and $300$ MeV.
For the quark matter densities considered here and a gap
of tens of MeV the GLBs mass is of the order of a few MeV.
We also note that in general $T_{\rm GLB}\simeq M_{\rm GLB}$.
In the following we define $E_{\gamma}$ as
 the photon energy from the ${\rm GLB}\rightarrow \gamma\gamma$ reaction.

\section{Scenario description and assumptions}

We begin with a hot quark star in the quark-gluon plasma 
phase  cooling and undergoing the transition into the 2SC phase.  
 As mentioned in the introduction,
 neutrino cooling will be dominant at first and the star
 rapidly reaches the temperature $T_{\rm c}\sim 0.57\Delta_{\rm 2SC}$
for the onset of the 2SC phase. The neutrino cooling is also  
dominant in the 2SC phase despite a decrease in neutrino emission 
 induced by pairing of quarks (Carter\&Reddy 2000; Page\&Usov 2002) until ultimately 
 the star cools to $T_{\rm GLB}$  at which point in time the GLBs start forming.
We assume for simplicity that the entire star
cools to $T_{\rm GLB}$ and any temperature
gradients in the early stages will be
smoothed out; which is not unreasonable given the extremely
large conductivity of the 2SC phase (e.g. Heiselberg\&Pethick 1993).
 All of the GLBs immediately decay into photons and   
 beyond this stage cooling is dominated by the photons
 which slowly diffuse out of the star.

\subsection{Total photon energy and plasma cut-off}

It was shown that 3/8 of the total number
 of gluons will form GLBs (Ouyed \& Sannino, 2001).
This implies that 3/8 of the nucleon binding energy, 938 MeV, 
is transformed to GLBs which amounts to roughly 
 350 MeV in photon energy per nucleon.
For a $10$ km radius quark star and 
matter densities on the order of $2\rho_{\rm N}$ ($\mu \sim 350$ MeV)
 this translates to a total energy
 $E_{\gamma, {\rm tot.}}\simeq 
 10^{53}$ ergs released as photons;  
$\rho_{\rm N}=2.71\times 10^{14}$ g cm$^{-3}$  is the nuclear
matter saturation density.
 However, let us recall that 
the emissivity of photons at energies below that of the
plasma frequency ($E_{\rm p}=\hbar\omega_{\rm p}\simeq$20-25~MeV
for the density regime representative of 2SC phase)
is strongly suppressed (Alcock, Farhi, \& Olinto 1986; Chmaj, Haensel,
\& Slomi\'nsli 1991; Usov 2001). 
As such, only the tail ($E_{\gamma} > 20-25$~MeV) of the 
distribution will exist inside the 2SC star. 
The true available total energy in photons  is  then 
$E_{\gamma, {\rm t}} \simeq \left [\int_{E_{\rm p}}^{\infty} f_{\gamma} dE_{\gamma}/\int_{0}^{\infty} f_{\gamma} dE_{\gamma}\right ] \times E_{\gamma, {\rm tot.}} \simeq 
 (10^{-6}$-$10^{-8})\times E_{\gamma, {\rm tot.}}\simeq (10^{45}$-$10^{47})$ ergs; 
here $f_{\gamma}$ is the Planck distribution; we note that different distributions
 lead to same values for $E_{\gamma, {\rm t}}$ but as we state in
\S 4.1 these photons will eventually acquire a Wien distribution as they leak out
 of the star.

\section{Cooling and lightcurves}

For the range in temperature, $T$,
considered here the vast majority of quarks in
 the star are in their ground states 
 $kT/E_{\rm F} << 1$ where $E_{\rm F}$
is the Fermi energy. 
The density of the free quarks 
 at temperature $T$ that will interact with the photons is 
\begin{equation}
N_q = \frac{\pi}{3} \left[  \frac{2e m_q k_B T}{\pi^2 \hbar^2}
   \right]^\frac{3}{2} ,
\label{qdeq}
\end{equation}
where $e$ is the electronic charge and $m_q$ is the quark mass.

The problem we are solving here is that of a uniform
sphere of constant density and constant temperature
gas of fermions and a homogeneous source of photons
with $E_{\gamma} > kT_{\rm q}$ where $T_{\rm q}$ is the
 the temperature of the quarks in this case.
The number of photons escaping as a function of time we calculated using 
the numerical method described by Sunyaev\&Titarchuk (1980, hereafter ST80;
see also Shapiro et al. 1976 and Miyamoto 1978).  The corresponding  
 diffusion equation in spherical coordinates is
\begin{equation}
\frac{\partial J}{\partial u} = \frac{1}{3\tau^2} \frac{\partial}{\partial \tau} \left( \tau^2 \frac{\partial J}{\partial \tau} \right)\ ,
\label{sta2}
\end{equation}
where $J$ is the average intensity of emission 
 and is a dimensionless  
quantity (following ST80 choice of units), 
$u$ is the dimensionless time ($u = c\sigma N_q t$; $c$ is the speed
 of light,   $t$ is the time in seconds, and $\sigma = \pi \lambda_{\rm q}^2$
where $\lambda_{\rm q} = \pi \left ( 2\pi\hbar c/\sqrt{(k_{\rm B}T)^2+ m_{\rm q}^2c^4} \right ) \simeq 1$ fm
is the quark wavelength\footnote{The cross-section $\sigma=\pi \lambda_q^2$ is
introduced to make use of dimensionless
 description  as explained in ST80 following
  Chapline\&Stevens (1973). We should note however that a complete
 mechanical prescription (using relativistic Compton cross-section in the appropriate regime
 where $h\nu > k T_{\rm q}$) was used in deriving  
 equation \ref{sta2}. More specifically the treatment of cross-section we
 adopted is that of Miyamoto (1978) which is consistent with the use of the Compton Focker-Planck equation.}),
 and $\tau$ is the dimensionless optical distance from the center of the star
 ($\tau = \sigma N_q r$ where $r$ is the distance from the center of the star).
 To find the escape time distribution
function $P(u)$ of a single photon we have to normalize
the function obtained $J(\tau_0,u)$ (see Appendix A in ST80 for details) so that
$\int_0^{\infty} P(u) du =1$, and
\begin{eqnarray}
P(u) &=& \frac{J(\tau_0,u)}{\int_0^{\infty}J(\tau_0,u) du}\nonumber\\
     &=& \frac{1}{\tau_0}\sqrt{\frac{3}{\pi u}} \left ( 1+2\frac{\sqrt{u}}{\tau_0}
\right) \exp\left (-\frac{\pi^2 u}{3\tau_0^2}\right )\ ,
\label{sta6}
\end{eqnarray}
where $\tau_0 = \sigma N_q R_q$ is estimated
at surface of the star of radius $R_q$.
Equation (\ref{sta6}) gives the normalized probability
of escape for photons from the surface of the star as 
a function of time. We then adjust the magnitude of the
distribution so that its area is equal to the total
number of photons, with $E_{\gamma} > E_{\rm p}$, from GLB decay.
The total number of photons escaping versus time was then converted
 to luminosity and to surface temperature, $T_{\rm s}$, using the black-body  formulation (see below). 

\begin{figure}
\includegraphics[width=0.5\textwidth]{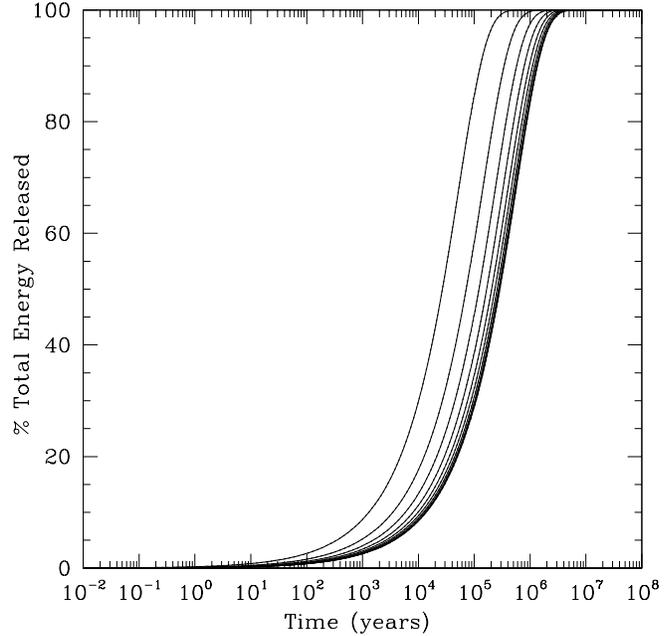}
\caption{The total energy that has escaped from the surface of a 10 km
radius quark star is plotted for increasing time.  The amount of escaped
 energy is shown as a percentage of the initial energy contained within the quark star at $t = 0$; the curves from left ($T_{\rm GLB}=1$ MeV)
 to right ($T_{\rm GLB}=10$ MeV) are in 1 MeV step.
} 
\label{fig:percentage}
\end{figure}

  Figure ~\ref{fig:percentage} shows the cumulative percentage of the total
 energy generated through GLB 
 decay that has been radiated from the surface of 10 km radius star 
as a function
 of time.  It takes at least  
$5\times 10^5$ years for the energy to escape the dense quark star and 
so we conclude that GLB decay in quark stars does not constitute
 a suitable inner
 engine for gamma ray bursts (GRBs) as originally 
claimed in Ouyed\&Sannino (2002); recent work by Ouyed et al.
(2005) suggests instead that quark stars in the  CFL phase constitute
better candidates for GRB inner engines. 
   However, Fig. 2 and Fig. 3 show that if 
 GLB decay does occur in a quark star the object will be extremely
 bright and should be readily observable.

\subsection{The escaping photons: {\it A Wien spectrum}} 

 In its simplest  approach the quark-photon interaction can be described
 using elementary random-walk theory which tells us that the average number
 of quark scatterings will be approximately equal
 to $\tau_{\rm s}^2$, with $\tau_{\rm s}= R/\lambda_{\nu}$ and 
 $\lambda_{\nu}\simeq 1$ fm being the Compton mean-free path of the photons.
 Furthermore, the average fractional energy transfer in
a quark-photon scattering will be of the order of $(2 kT_{\rm q}/m_{\rm q}c^2)$.
 For $\tau_{\rm s} >> 1$, as it is in our case, 
  one can show that the initial photon spectrum is distorted 
 into a distribution having approximately the shape of a Wien spectrum
 (e.g., Chapline \& Stevens 1973). In general, it is
found that an equilibrium solution to the
 Compton Fokker-Planck equation is a Bose-Einstein distribution which,
 for $h \nu > k T_{\rm q}$ and large $\tau$, approaches a Wien spectrum
 (Miyamoto 1976 and references therein).
That is, by the time the photon escape the star
 their distribution will be  given by 
 $F_{\nu} \propto \nu^3  \exp(-h\nu / k T)$.

\subsection{Thermalization: {\it A black-body spectrum}}

A property of quark stars is the existence
 of an ``electrosphere" extending a few hundred to a thousand Fermi above the
 surface of the star (see Usov, Harko \& Cheng for a recent study)\footnote{A charge neutralizing surface electron layer
  can be realized in the 2SC, the gapless CFL (gCFL)
 or crystalline color superconductive phases (Alford\&Rajagopal 2002; Shovkovy\&Huang
2003), but
 not in the pure CFL phase (see however Usov 2004).}.  Based on a collisional
treatment of photons on single electrons and  given the
  electron density in the electrosphere,  $n_{e}\sim 10^{-5}-10^{-4}$ fm$^{-3}$,
 one can show that photons with energies $E_{\gamma} > 6.5 MeV$ will
 not be affected by the layer (Cheng \& Harko 2003; Jaikumar et al. 2004). In other words, the escaping photons do not scatter often in the electropshere to become thermalized.
 
However, the outflowing high energy ($E_{\gamma} >> m_{\rm e}c^2$)
 photons enter a region rich in $(e^+e^-)$ pairs
 and low energy ($<< E_{\gamma}$) photons (Aksenov et al. 2004). 
 The corresponding `compactness parameter'  
  $l =L_{\gamma}\sigma_{\rm T}/4\pi R c^2 >> 1$ (e.g. Frank,
King, \& Raine 1992) implies ideal conditions
 for $\gamma\gamma\rightarrow e^+e^-$ processes to take place
  quickly leading to thermal equilibrium (e.g. Paczy\'nski 1990).
 We thus expect the Wien distribution of the escaping
 photons to evolve into a black-body although
  a residual Wien tail in gamma-rays ($\ge 20$-$25$ MeV)
   could survive as a signature of the original distribution.

\subsection{Crust effects}

The electrosphere allow for radial electric fields whose
 magnitude can support matter against the star's gravity. 
 The mass of the crust which can exist suspended by this electric field  
 cannot exceed $10^{-5}M_{\odot}$ (Alcock et al. 1986; Horvath
et al. 1991). Its effect can be quantified by
relating  the ``internal'' temperature $T_{\rm s}$ or the temperature at the bottom 
of the crust to the effective temperature at the surface of the crust via the
so-called Tsuruta law (Tsuruta 1979).
The formula for thick crusts 
 is (see also Shapiro\&Teukolsky 1983, p330)
$T_{\rm e}= (10 T_{\rm s})^{2/3}$.
We note however that for quark stars the crust is much thinner
than in the NSs.
 Therefore, we will investigate the two extreme cases
of negligible or no crust ($T_{\rm e}\simeq T_{\rm s}$;
e.g. Pizzochero 1991) and thick
or maximum crust ($M_{\rm max.}\simeq 10^{-5}M_{\odot}$) using the Tsuruta formula. 
In both cases, the effective temperature as seen by a distant observer
is $T_{\rm e}^{\infty}=T_{\rm e}\sqrt{1-R_{\rm Sch.}/R}$
where $R_{\rm Sch.}$ is the star's Schwarschild radius.

\subsection{Comparison to observations}

\begin{figure*}
\includegraphics[width=\textwidth]{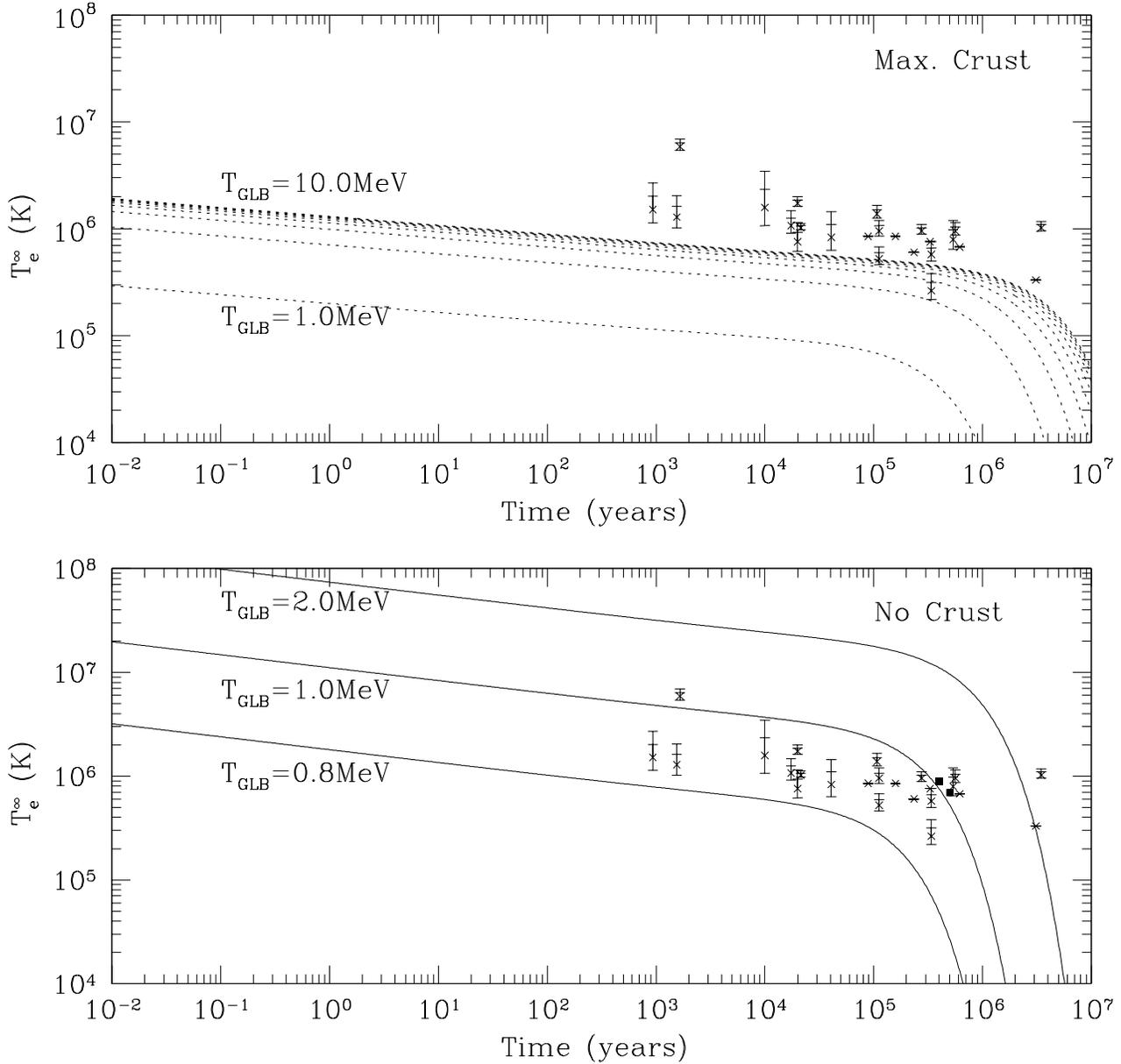}
\caption{The time evolution of the  
 temperature, $T_{\rm e}^{\infty}$ 
  (as seen by a distant observer) for a star of a $10$ km radius.
Shown are cooling curves for different
initial temperatures $T_{\rm GLB}$. Solid lines
are for a star with no crust
while the dotted ones is for the maximum crust case. 
The cooling curves begin approximately 2 weeks after $t$ = 0.
Experimental data taken by {\it ROSAT} and {\it ASCA} 
(Table 7 in Weber 2005) are also shown. 
 }
\label{fig:cooling}
\end{figure*}

The related cooling curves are 
shown in Figure~\ref{fig:cooling} 
 for different initial temperatures $T_{\rm 0}$ and
for a star of radius 10 km.
Models with maximum crusts do not appear to be successful when
 compared to cooling of neutron stars if these are quark stars.
The corresponding cooling curves start overlapping with observed data
 for  $T_{\rm GLB} > 0.57 \Delta_{\rm 2SC}$ which is outside the
2SC phase where GLBs cannot form. 
Such temperatures will not be sustained since we
 expect neutrinos will drive $T_{\rm GLB}$ to lower values early
 in the star's cooling history.
 When compared to observed data our cooling curves agree best with  
 models of stars with thin or no crusts and  for temperatures 
$0.8 < T_{\rm GLB} < 2$ MeV. The regime favored by the best fits correspond to 
 densities $\rho_{\rm 2SC}=(2.5$-$3.0)\times
\rho_{\rm N}$ (or $\mu= 350$-$450$ MeV) and energy gaps 
$\Delta_{\rm 2SC} = 15$-$30$ MeV (\S 2). 

In general, in our model best fits to observations favor quark stars with thin 
or no crusts. This is also consistent with the fact that quark stars
 are expected to be born bare given the extreme temperatures involved
 during their formation.  In the so-called `Quark-Nova' scenario
 (Ouyed et al. 2002; Ker\"anen et al. 2005) for example the crust
 could be regenerated from fall-back material but it is expected to be
 tiny.  This brings to the special family of hot radio-quiet isolated neutron stars
 (INSs) within the picture we presented so far. Specifics below.

\section{Radio-quiet isolated neutron stars} 

\subsection{Observations}

 There are about 7 radio-quiet isolated neutron stars (INSs) that show features
 that might be of interest to us. 
There are important clues from observations that are worth mentioning:
 (i) While recent XMM-Newton observations of 4 
 radio-quiet INSs show clear features (in particular RX J0720.4-3125 
 show  a clear broad feature at around 310 eV; Haberl et al. 2004), RX J1856.5-3754 remains featureless; (ii) it is believed that the observed broad features
 are absorption due to the proton cyclotron line. These hydrogen lines it
 was suggested could arise from a pure hydrogen atmosphere (van Kerkwijk
 et al. 2004); (iii) what is further  remarkable about these objects is the
 optical excess observed in their spectra as compared to the
 X-ray component (e.g. kaplan et al. 2003). Recent studies of this
 phenomenon show that a thin (1 cm thin and not exceeding $10^{14}$ g)
 layer of hydrogen atmosphere can well represent both the optical and 
 X-ray data without invoking additional thermal component (Motch et al. 2003);
 (iv) finally, there are indications that accretion might not
 be important since the rates must be many orders of magnitudes
 smaller than the Bondi-Hoyle rates of $\sim 10^8$ g/s (Bondi \& Hoyle 1944; Bondi 1952). Motch et al. (2003) estimate from fits to optical emission
 that the total mass of hydrogen is no more than $2\times 10^{12}$ g which
 translates into accretion rates not exceeding $1$ g/s. A second clue that 
 accretion might not be at play in these sources is the
$H_{\alpha}$ nebula surrounding $RX J1856.5-3754$ which can be attributed
 to a relativistic wind (van Kerkwijk \& Kulkarni 2001).

\subsection{Our scenario}

The newly born bare quark star could regain its crust
  by accretion from the ISM or from fall-back material
 during its formation (Ker\"anen et al. 2005). In both of theses possibilities  
  a tiny hydrogen envelope can only be explained  
 if accretion rates are less than $1$ g/s. Such rates are puzzling
 and difficult to account for and we would
 rather assume that accretion is irrelevant (see point (iii) in \S 5.1)
  and that the star regained a crust
 from fall-back material as in the case of the `Quark-Nova' scenario (Ouyed et
 al. 2002; Kera\"anen et al. 2005) or any similar formation scenarios of quark stars.
 We should however mention the scenario  discussed in Usov (1997) where
 it is argued  that in the case of more or less spherical (Bondi) accretion onto a
nearly bare quark star only a small part of falling gas is collected at
the stellar surface. The bulk of this gas passes through the quark
surface leaving a rather low-mass steady hydrogen atmosphere ($\sim 
10^{12}$ g).  Below we discuss another possibility  whereas
 the original crust is slowly depleted in time that might
  possibly explain the clustering in time and temperature of INSs.

An important intrinsic property of quark stars is the dependency
 of the size of the crust on the properties of the electropshere (Alcock et al.
 1986; Usov et al. 2005) which in turn   
 are very sensitive to the global charge neutrality of the star and its
 temperature. While in the case of pure CFL
 it has been argued in the literature that in such a state  
  the star is globally neutral by construction (Rajagopal \& Wilczek 2001;
 see however Usov 2005) we ask ourselves the question
 of what happens in the case of 2SC, g2SC and gCFL? As these
 stars cool one could imagine that they slowly 
 evolve to a more globally neutral configuration.
  Presumably the electrosphere and the corresponding electic
field are reduced in time  
 which would correspond to the crust shrinking to eventually reach
   infinitesimal size when most of the electron in the electropshere have
 been reabsorbed by the quark matter. What is interesting in this
 scenario we suggest - which remains to be confirmed -
 is the fact that the bottom of the
 crust (dominated by the settled heavy elements) sinks closer
 to the quark matter surface. 
Over time the crust material is slowly
 depleted/deconfined  from the bottom leaving
 the lighter upper atmosphere material. 

\begin{figure}
\centering
\includegraphics[width=0.5\textwidth]{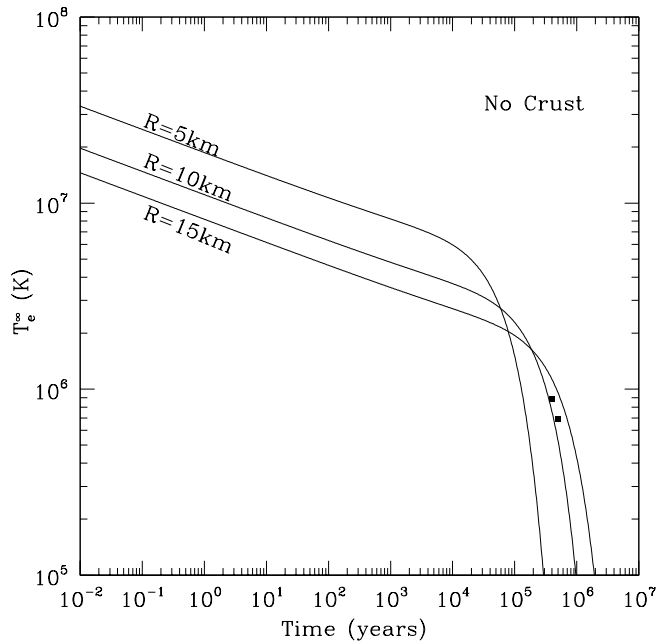}
\caption{The time evolution of the  
 temperature, $T_{\rm e}^{\infty}$ 
  (as seen by a distant observer) for $T_{\rm GLB}=1$ MeV, and
 for different star radius $R$. 
The cooling curves begin approximately 2 weeks after $t$ = 0.
The observed data are RX J185635-3754 (lower 
square) and RX
J0720.4-3125 (upper square).
 }
\label{fig:cooling_rad}
\end{figure}

In our scenario then the smaller the crust gets as the star cools and reduces its  
 electrosphere the richer in lighter elements (and ultimately in hydrogen) it becomes.  
The end result we suggest is a tiny envelope reminiscent of the thin
 hydrogen envelope inferred from observations. Hence we can define
 a universal critical temperature $T_{\rm H}$ below which the envelope
 becomes transparent leading to the optical excess.   The rate of depletion
 of the envelope closely follows the cooling rate/curve. As such,
 the size of the envelope will drop quicker as the stars
 enter the faster cooling beyond the $10^{5}$-$10^{6}$ years epoch.   
  This one-to-one relationship could in principle account
 for the  observed clustering in time and temperature of the 7 radio-quiet INSs as well as their their
 associated optical excess.
The critical temperature $T_{\rm H}$ among others features
 inherent to our model and the observational consequences 
 will be explored in more details in an upcoming paper.
  For now we are tempted to speculate that radio-quiet INSs  
are quark stars\footnote{The  possibility that RX J1856.5-3754 is
 a bare quark star has been put forward in the literature (Xu 2003; Turolla et al. 2004).
 The novelty in our model resides
 in the GLB decay as an extra energy source. It has also been suggested
 that this object might be a naked neutron star where cohesive effects 
 induced by moderately strong magnetic field leads to condensation in the
 atmosphere and the surface external layers 
  (Perez-Azorin et al.  2005; van Adelsberg et al. 2004).} in the superconductive phase cooling through the slow release of photons from GLB decay or similar processes. 
  For completeness, shown in Figure \ref{fig:cooling_rad} are RX J185635-3754 
 and RX J0720.4-3125 
(Haberl et al. 1997; Walter\&Lattimer 2002; Kaplan et al. 2003a; Walter et al. 2004) which line up closely with the $T_{\rm GLB}=1$ MeV curve for a 10 km quark star.

\section{Conclusion}

In this work we investigated the effect of photons from GLB decay 
on the  cooling history of quarks stars in the 2SC phase.
  While the early cooling history is dominated by neutrino
 emission,  we find that further cooling is driven by
 photons, from the GLB decay, slowly diffusing out of the star.
 This cooling mechanism could be potentially important
 in situations or cases where the  neutrino emission is heavily 
suppressed by pairing of quarks.  Up to $10^{45}-10^{47}$ erg of photon
energy can be provided by the GLB decay radiated away by the star
 over millions of years. The persistent blackbody emission
 offers a different picture from other models of cooling
of quark stars where the photon spectrum is predicted to vary in time.
 Our fit to cooling data of observed INSs suggests that these are quark stars
with extremely thin or no crust where GLB decay has taken effect.
  The fits favor gap energy $\Delta_{\rm 2SC} = 15$-$35$ MeV
 with corresponding densities $\rho_{\rm 2SC}=(2.5$-$3.0)\times 
\rho_{\rm N}$;  this might be of relevance to Quantum-Chromodynamics.
Let us state again that in principle our picture  
applies to quark stars where photon generation mechanisms are at
play.  If correct, our model combined with more observations
of cooling of isolated compact stars, in particular
 hot radio-quiet ones, could be used to
constrain other fundamental parameters of quark matter and its exotic phases.

\section*{acknowledgements}
R.O. thanks P. Jaikumar, R. Rapp, D. Leahy, K. Mori, V. Usov, and M. van Kerkwijk for helpful comments.
 The authors thank the Nordic Institute
for Theoretical Astrophysics (NORDITA), where this
work was initiated, for its hospitality.
The research of R.O. is supported by an
operating grants from the Natural Science and Engineering Research
Council of Canada (NSERC) as well as the Alberta Ingenuity Fund (AIF).

\label{lastpage}


\begin{thebibliography}{99}


\bibitem[]{Aguilera05} Aguilera, D. N., Blaschke, D., \& Grigorian, H. 2005,
 Nucl. Phys. A, 757, 527

\bibitem[]{Aksenov04} Aksenov, A. G., Milgrom, M., \& Usov, V. V. 2004,
 609, 363

\bibitem[]{Alcock86} Alcock, C., Farhi, E., \& Olinto, A. 1986,
 ApJ, 310, 261

\bibitem[]{Alford02} Alford, M., \& Rajagopal, K. 2002, High Ener. Phys. 06, 31

\bibitem[]{Alford04} Alford, M., Kouvaris, Ch., \& Rajagopal, K. 2004 [astro-ph/0407257]

\bibitem[]{Blaschke00} Blaschke, D., Kl\"ahn, T., \& Voskresensky, D. N. 2000, ApJ,
533, 406

\bibitem[]{Blaschke01} Blaschke, D., Grigorian, H., \& Voskresensky, D. N. 2001,
A\&A, 368, 561

\bibitem[]{Blaschke05} Blaschke, D., Grigorian, H., Khalatyan, A., \& Voskresensky, D. N. 2005, Nucl. Phys. B (Proc. Suppl.), 141, 137

\bibitem[]{bondi44} Bondi, H., \& Hoyle, F. 1944, MNRAS, 104, 421

\bibitem[]{bondi52} Bondi, H. 1952, MNRAS, 112, 195

\bibitem[]{carter00} Carter, G. W., \& Reddy, S. 2000, Phys. Rev. D, 62, 103002

\bibitem[]{Casalbuoni03} Casalbuoni, R. [hep-ph/0310067]

\bibitem[]{Casalbuoni05} Casalbuoni, R., Gatto, R., Mannarelli, M., Nardulli,
 G., \& Ruggieri, M. 2005, Phys. Let. B, 605, 362

\bibitem[]{Chapline73} Chapline, G. Jr., \& Stevens, J. 1973, ApJ, 194, 1041

\bibitem[]{Cheng03} Cheng, K. S., \& Harko, T. 2003, ApJ, 596, 451

\bibitem[]{Chmaj91} Chmaj, T., Haensel, P., \& Slomi\'nski, W. 1991, Nucl. Phys. B, 24, 40

\bibitem[]{Frank92} Frank, J., King, A., \& Raine, D. 1992,
 `Accretion Power in Astrophysics' (Cambridge University Press)

\bibitem[]{haberl97} Haberl, F. et al. 1997, A\&A, 326, 662 

\bibitem[]{haberl04} Haberl, F. et al. 2004, A\&A, 419, 1077 

\bibitem[]{He05} He, L., Jin, M., \& Zhuang, P., 2005, Phys. Rev. D, submitted [hep-ph/0505061]

\bibitem[]{Heiselberg93} Heiselberg, H., \& Pethick, C. J. 1993,
Phys. Rev. D, 48, 2916

\bibitem[]{Horvath91} Horvath, J. E., Benvenuto, O. G., \& Vucetich, H. 1991,
Phys. Rev. D, 44, 3797

\bibitem[]{Huang05} Huang, M., \& Shovkovy, I. A. 2005, Phys. Rev. D, 70, 094030

\bibitem[]{Jaikumar04} Jaikumar, P., Gale, Ch., Page, D., \& Prakash,
 M. 2004, Phys. Rev. D 70, 023004

\bibitem[]{Juett02} Juett, A. M., Marshall, H. L., Chakrabarty, D., \& Schulz, N. S. 2002, ApJ, 568, L31

\bibitem[]{kaplan03a} Kaplan, D. L. et al. 2003a, ApJ, 588, L33

\bibitem[]{kaplan03b} Kaplan, D. L. et al. 2003b, ApJ, 590, 1008

\bibitem[]{keranen05} Ker\"anen, P., Ouyed, R., \& Jaikumar, P. 2005,
 ApJ, 618, 485

\bibitem[]{miyamoto78} Miyamoto, S. 1978, A\&A, 63, 69

\bibitem[]{motch03} Motch, C., Zavlin, V. E., \& Haberl, F. 2003, A\&A, 408, 323

\bibitem[]{Ouyed01} Ouyed, R., \& Sannino, F. 2001, Phys. Lett. B. 511, 66

\bibitem[]{OuyedDey} Ouyed, R., Dey. J., \& Dey, M. 2002, A\&A, 390, L39

\bibitem[]{Ouyed02a} Ouyed, R. 2002, in Compact Stars in the QCD Phase
Diagram, Proceedings, Copenhangen, August, 2001, eds  R. Ouyed and F. Sannino., p.209 (www.slac.stanford.edu/econf/C010815), astro-ph/0201408

\bibitem[]{Ouyed02b} Ouyed, R., \& Sannino, F. 2002, A\&A, 387, 725

\bibitem[]{Ouyed04} Ouyed, R., Elgar{\o}y, {\O}, Dahle, H., \& Ker\"anen, P.
A\&A, 2004, 420, 1025

\bibitem[]{Ouyed05} Ouyed, R., Rapp, R., \& Vogt, C. 2005, ApJ, 632, 1001

\bibitem[]{Paczynski90} Paczy\'nski, B. 1990, ApJ, 363, 218

\bibitem[]{page02} Page, D., \& Usov, V. V. 2002, Phys. Rev. Lett., 89, 131101

\bibitem[]{Patel03} Patel, S. K. 2003, ApJ, 587, 367

\bibitem[]{Perez05} Perez-Azorin, J. F., Miralles, J. A., \& Pons, J. A. 2005, A\&A, 433, 275

\bibitem[]{Pizzochero91} Pizzochero, P. M. 1991, Phys. Rev. Lett., 66, 2425

\bibitem[]{Rajagopal01} Rajagopal, K. \& Wilczek, F. 2001, Phys. Rev. Lett., 86, 3492

\bibitem[]{Rischke01} Rischke, D. H., Son, D. T., \& Stephanov, M. A. 2001, Phys. Rev.
Lett. 87, 062001

\bibitem[]{Sannino02} Sannino, F., Marchal, N., \& Sch\"afer, W. 2002, Phys. Rev. D 66

\bibitem[]{Schaab96} Schaab, C., Weber, F., Weigel, M. K. \& Glendenning, N. K.
1996, Nucl. Phys. A, 605, 531

\bibitem[]{Schaab00} Schaab, C., Hermann, B., Weber, F., \& Manfred, K.
2000, ApJ, 480, L111

\bibitem[]{Shapiro76} Shapiro, S. L., Lightman, A., \& Eardley, D. 1976, ApJ, 204, 187

\bibitem[]{STbook83} Shapiro, S. L. \& Teukolsky, S. A 1983, {\it Black Holes, White Dwarfs and Neutron Stars}, John Wiley \& Sons, Inc. N. Y.

\bibitem[]{Shovkovy03} Shovkovy, I. A., \& Huang, M. 2003,  Phys. Lett. B, 564, 205 

\bibitem[]{Shovkovy04} Shovkovy, I. A., R\"uster, S. B., \& Rischke, D. H. 2004 [astro-ph/0411040]

\bibitem[] {sunyaev80} Sunyaev, R. A. \& Titarchuk, L. G. 1980, 
A\&A, 86,121.

\bibitem[1979]{tsuruta79} Tsuruta, S. 1979, Phys. Rep., 56, 237

\bibitem[2004]{turolla04} Turolla, R., Zane, S. \& Drake, J. J. 2004, ApJ, 603, 265

\bibitem[]{Usov1997} Usov, V. V. 1997, ApJ, 481, L107

\bibitem[]{Usov01} Usov, V. V. 2001, ApJ, 550, L179

\bibitem[]{Usov04} Usov, V. V. 2004, Phys. Rev D 70, 067301  

\bibitem[]{Usov05} Usov, V. V., Harko, T., \& Cheng, K. S. 2005, ApJ, 620, 915

\bibitem[]{Adelsberg04} van Adelsberg, M., Lai, D., \& Potekhin, A. Y. 2004 [astro-ph/0406001]

\bibitem[]{marten04b} van Kerkwijk, M. H., \& Kulkarni,
 S. R. 2001, A\&A, 380, 221

\bibitem[]{marten04} van Kerkwijk, M. H., Kaplan, D. L., Durant, M., Kulkarni,
 S. R., \& Paerels, F. 2004, ApJ, 608, 432

\bibitem[]{Vogt04} Vogt, C., Rapp, R., \& Ouyed, R. 2004, Nucl. Phys. A, 735, 543

\bibitem[]{walter02} Walter, F. M., \& Lattimer, J. M. 2002, ApJ, 576, L145

\bibitem[]{Weber05} Weber, F. 2005, Prog. Part. Nucl. Phys. 54, 193 

\bibitem[]{Xu03} Xu, R. X. 2003, ApJ, 596, L59


\end{thebibliography}
\end{document}